\documentclass[a4paper]{article}

\usepackage{icrc2013}
\usepackage{xspace}
\usepackage{nth}
\usepackage{textcomp}
\usepackage{amssymb}
\usepackage[english]{babel}

\newcommand{\pe}{p.e.\xspace{}}

\title{FACT - How stable are the silicon photon detectors?}

\shorttitle{FACT - How stable are the silicon photon detectors?}

\newcommand{\ethz}{$^1$}
\newcommand{\tudo}{$^2$}
\newcommand{\uniw}{$^3$}
\newcommand{\epfl}{$^4$}
\newcommand{\unige}{$^5$}

\authors{
T.~Bretz\ethz,
A.~Biland\ethz,
J.~Bu\ss\tudo,
D.~Dorner\uniw,
S.~Einecke\tudo,
D.~Eisenacher\uniw,
D.~Hildebrand\ethz,
M.~L.~Knoetig\ethz,
T.~Kr\"ahenb\"uhl\ethz,
W.~Lustermann\ethz,
K.~Mannheim\uniw,
K.~Meier\uniw,
D.~Neise\tudo,
\mbox{A.-K.}~Overkemping\tudo,
A.~Paravac\uniw,
F.~Pauss\ethz,
W.~Rhode\tudo,
M.~Ribordy\epfl,
T.~Steinbring\uniw,
F.~Temme\tudo,
J.~Thaele\tudo,
P.~Vogler\ethz,
R.~Walter\unige,
Q.~Weitzel\ethz,
M.~Z\"anglein\uniw $\;\;$
(FACT Collaboration)
}
\afiliations{
\ethz ETH Zurich, Switzerland $\;\;-\;\;$
   Institute for Particle Physics, Schafmattstr.~20, 8093 Zurich\\
\tudo Technische Universit\"at Dortmund, Germany $\;\;-\;\;$
   Experimental Physics 5, Otto-Hahn-Str.~4, 44221 Dortmund\\
\uniw Universit\"at W\"urzburg, Germany $\;\;-\;\;$
   Institute for Theoretical Physics and Astrophysics,
   Emil-Fischer-Str.~31, 97074 W\"urzburg,\\
\epfl EPF Lausanne, Switzerland  $\;\;-\;\;$
   Laboratory for High Energy Physics, 1015 Lausanne\\
\unige University of Geneva, Switzerland $\;\;-\;\;$
   ISDC Data Center for Astrophysics, Chemin d'Ecogia~16, 1290 Versoix \\
}

\email{thomas.bretz@phys.ethz.ch}

\abstract{
The First G-APD Cherenkov telescope (FACT) is the first telescope using
silicon photon detectors (G-APD aka. SiPM). The use of Silicon devices
promise a higher photon detection efficiency, more robustness and
higher precision than photo-multiplier tubes. Since the properties of
G-APDs depend on auxiliary parameters like temperature, a feedback
system adapting the applied voltage accordingly is mandatory.

In this presentation, the feedback system, developed and in operation
for FACT, is presented. Using the extraction of a single
photon-equivalent (pe) spectrum as a reference, it can be proven that
the sensors can be operated with very high precision. The extraction of
the single-pe, its spectrum up to 10\,pe, its properties and their
precision, as well as their long-term behavior during operation are
discussed. As a by product a single pulse template is obtained. It is
shown that with the presented method, an additional external
calibration device can be omitted. The presented method is essential
for the application of G-APDs in future projects in Cherenkov astronomy
and is supposed to result in a more stable and precise operation than
possible with photo-multiplier tubes.
}


\begin{document}
\maketitle

\renewcommand\textfloatsep{1em} 

\section{Introduction}

The First G-APD Cherenkov Telscope (FACT,~\cite{bib:design}) is the first
installation of a complete focal plane using silicon
photo sensors. It is in operation since Oct.\ 2011
and its main goals are to prove the applicability
of Geiger-mode avalanche photo diodes (G-APD) 
for focal planes with changing environmental 
conditions and the long-term monitoring of
the brightest TeV blazars.

While G-APDs are very robust and easy in their application
taken their relatively low bias voltage of usually less than 100\,V
into account, their operation properties are also very sensitive
to their temperature and the applied voltage.

Since active cooling and the achievement of a homogeneous
temperature on a large surface exposed to the environment is
difficult, the camera only features a passive cooling and thermal
insulation to ensure that the waste heat of the electronics does not
significantly heat the sensors.

By thermal design, the temperature gradient in the focal plane
is small enough to allow powering of four and five sensors
at the same time with the same bias voltage. They have been 
sorted accordingly.
The bias voltage supply system comprises 320 channels in total
and allows to adapt the voltage with a precision of about 
22\,mV. Simultaneously, it enables the readout of the
provided current with a precision of 1.2\,\(\mu A\).

\section{The feedback system}

To keep the gain, optical crosstalk probability and afterpulse probability
of G-APDs stable, a constant overvoltage must be applied. The applied
overvoltage is the voltage difference between the applied absolute voltage
and the breakdown voltage, which is a function of temperature.
The dependence of the breakdown voltage on temperature is linear, well defined and
identical for all channels and can thus be corrected adapting
the applied voltage according to the measured temperature.

To correct for the change in overvoltage, induced by the change in current and
temperature, a feedback system is applied. As feedback values,
31 temperature sensors in the sensor compartment are available,
as well as the current readout of each bias voltage channel. 

The bias voltage is distributed to the G-APDs using a passive filter
network. Its serial resistance induces a voltage drop depending on
the current flowing. 
Since bright ambient light condition as moon lit nights can induce 
count rates up to 1000 times higher than during dark night conditions,
every change in ambient light level would directly influence the applied
overvoltage. The voltage drop at the
G-APD can be calculated from the measured current and resistances.

For the correction of both effects, a feedback loop has been implemented
in the slow control software, which calculates the correction offset
according to the feedback values.
Since temperature changes are assumed to be slow, the temperature
is evaluated every 15\,s. Changes in current can be comparably fast due
to bright stars moving over the field-of-view, therefore, the voltage 
drop is calculated once every second. 

\section{Measuring the stability}

To measure the performance of the feedback system, the G-APD
properties are evaluated. Different methods 
to determine the dependence on the applied bias supply voltage, the temperature
and the ambient light condition, have been applied and are discussed
hereafter.

\subsection{Light pulser measurements}

As an indirect measurement of the gain, the measured  
amplitude of a light pulse is used. The light pulses are emitted by a
light pulser installed in the reflector
dish.  Since its light yield is temperature stabilized, the measured
signal amplitude is an indirect measurement of the gain of all channels.

For each measurement, the light-pulser is flashed one thousand times
and the readout is self-triggered by the camera's trigger system.
The signal is extracted
either by searching the peak or integrating the pulse over a certain
range around the peak, both done by a \nth{3} order spline
interpolation. For each channel, the average amplitude is calculated
and the median of all channel is taken, hereafter, just called
{\it light-pulser amplitude}.

From random triggered events, a measure for the ambient light
level can be derived calculating the fluctuations of the recorded
signal. For this, the signal-extraction algorithms is applied in
each event at a random position. The strength of the fluctuation
corresponds to the background photon rate. 
To be able to correlate the measured values with temperature,
an average of all temperature sensors is calculated during these five minutes.

\begin{figure}[t]
\begin{center}                
 \includegraphics*[width=0.48\textwidth,angle=0,clip,trim=0 0 1.7cm 1.3cm]{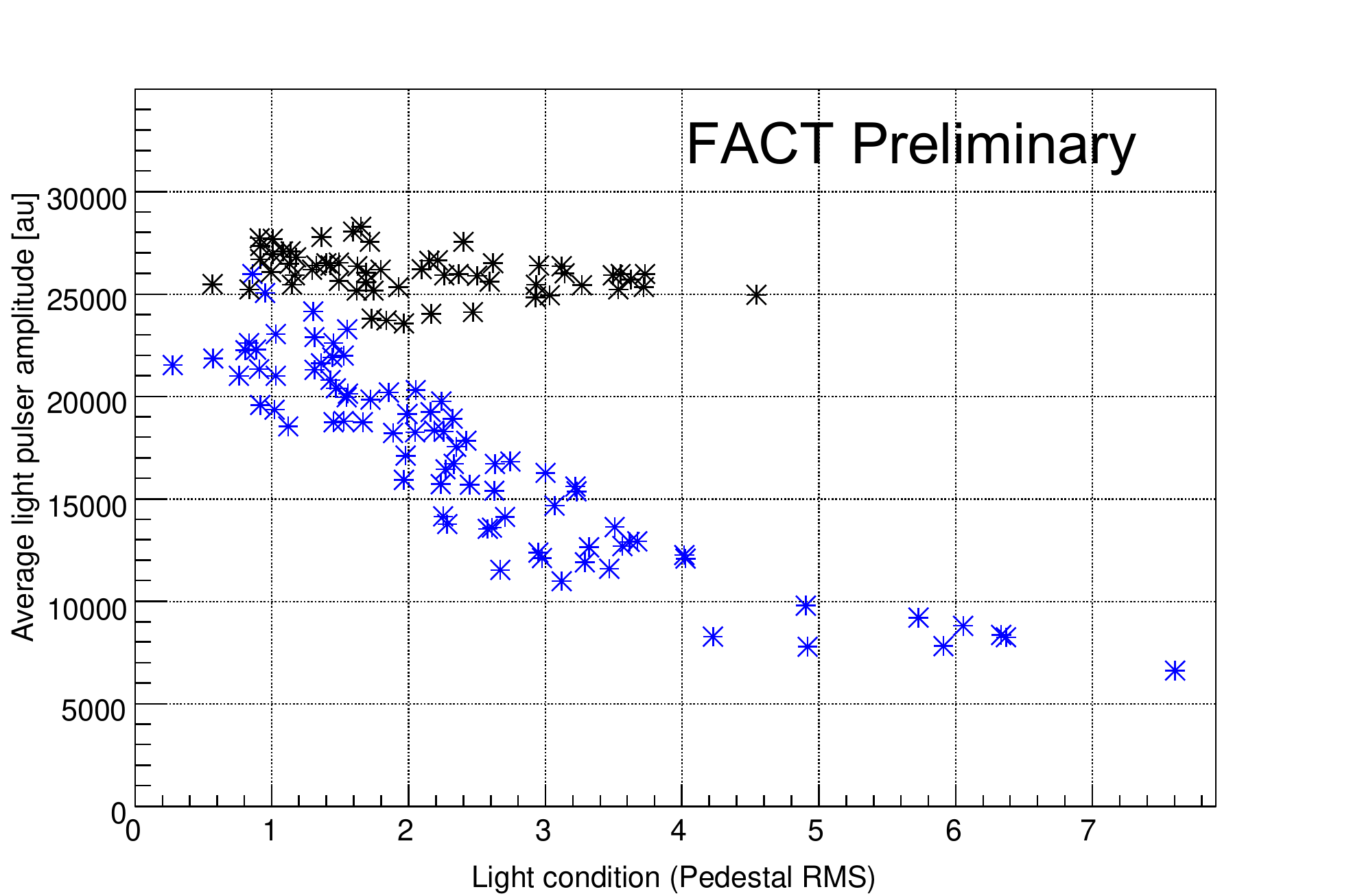}
\caption{The median of the average extracted light-pulser signal 
versus the measured pedestal rms. The blue asterix' correspond
to data taken with only the temperature feedback switched on,
the black dots have been taken with the full current feedback enabled.}
\label{fig:AmplVsPed}
\end{center}
\end{figure}

The measured light-pulser amplitude is shown in Fig.~\ref{fig:AmplVsPed}
versus average temperature with only the temperature-feedback switched on (blue)
and with the whole feedback system enabled (black). 
With just the temperature correction, a clear
dependence of the light-pulser amplitude on the ambient light level is
visible, as expected.
Applying the correction based on the measured current fully
corrects this dependency.

\begin{figure}[htb]
\begin{center}                
 \includegraphics*[width=0.48\textwidth,angle=0,clip]{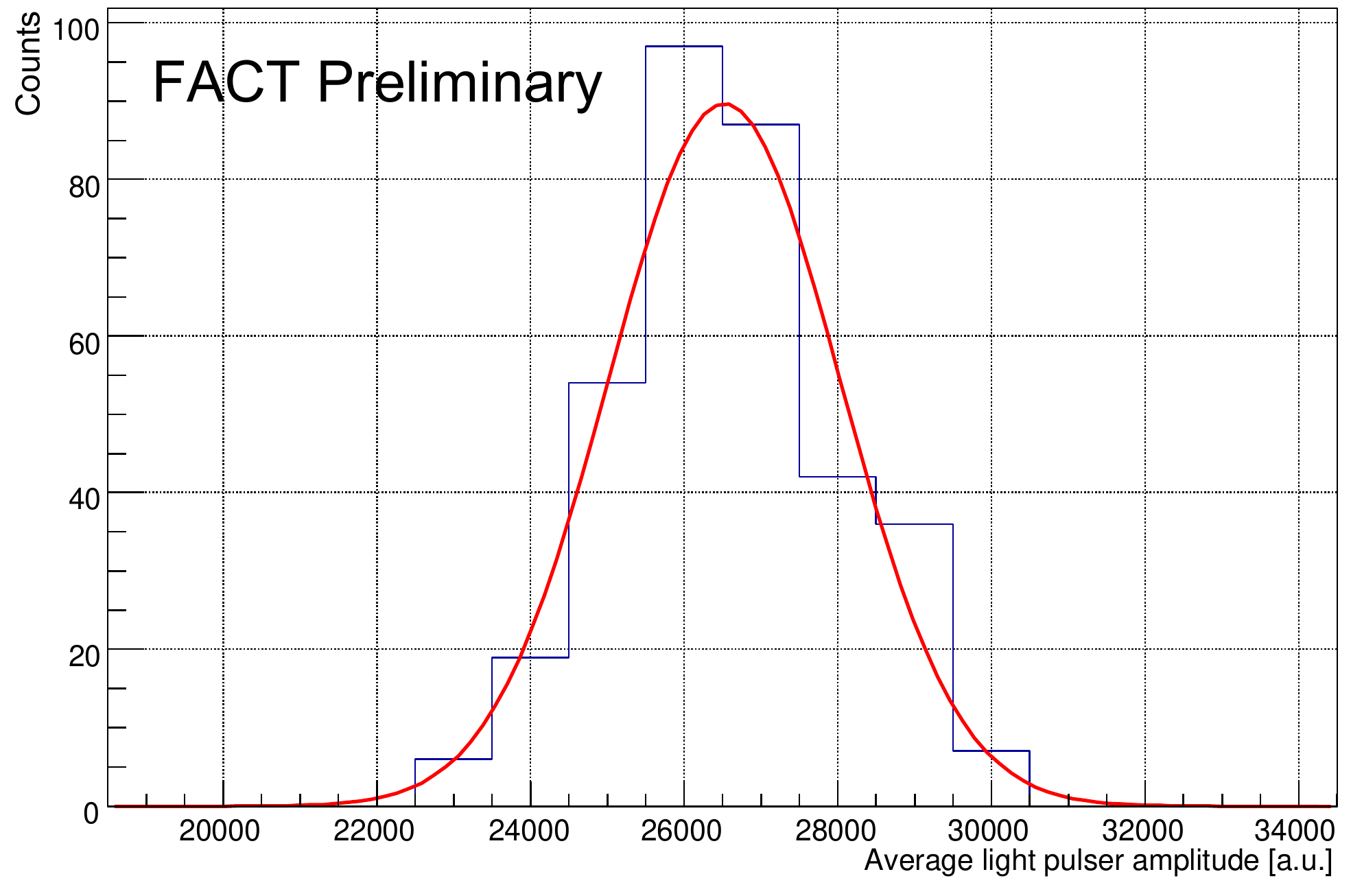}
\caption{Distribution of light-pulser amplitudes measured at average
compartment temperatures between 5\textdegree{}C and 25\textdegree{}C
fitted with a Gaussian.}
\label{fig:TempDist}
\end{center}
\end{figure}

\begin{figure*}[htb]
\begin{center}
 \includegraphics*[width=0.48\textwidth,angle=0,clip]{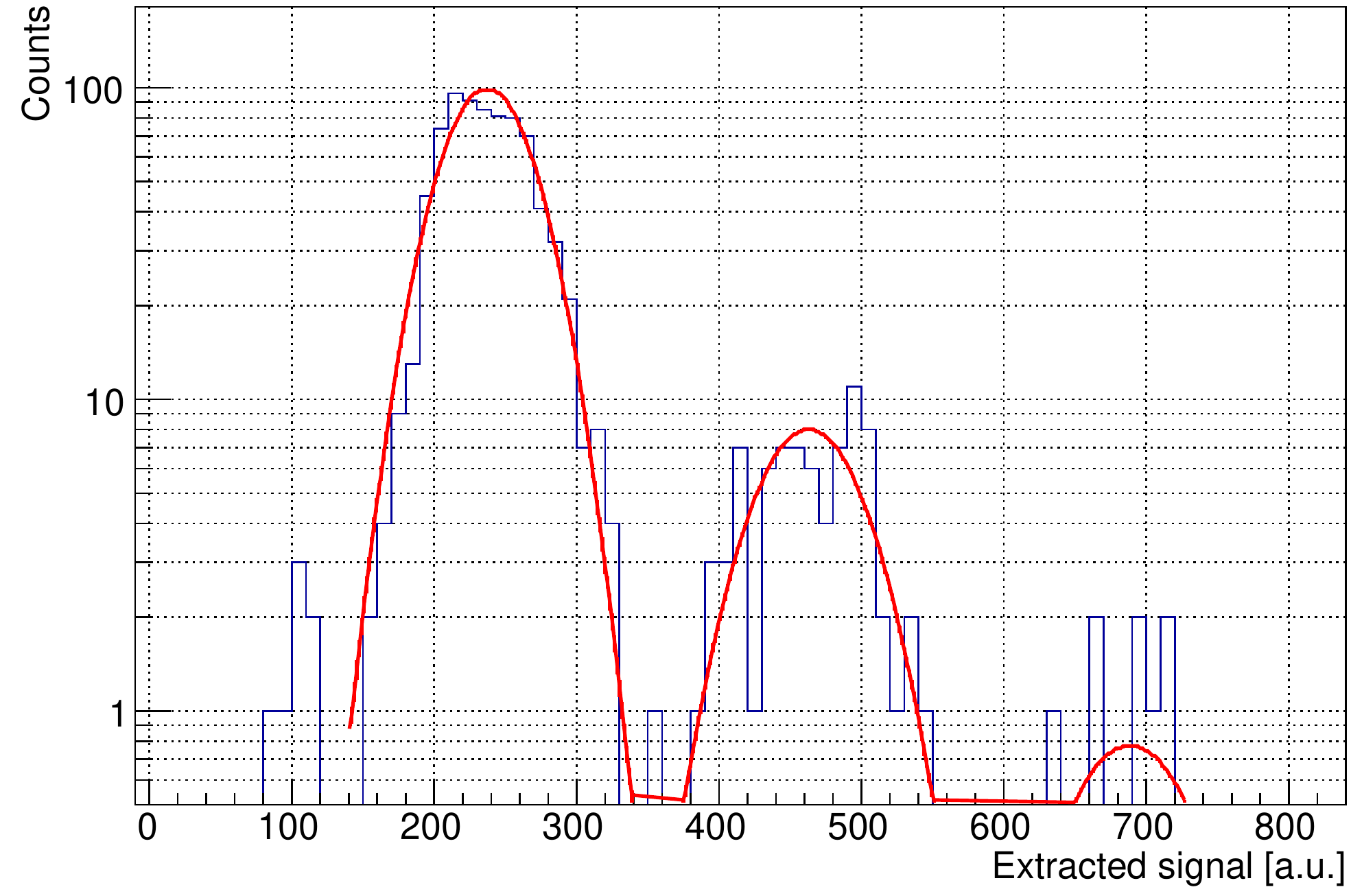}
 \includegraphics*[width=0.48\textwidth,angle=0,clip]{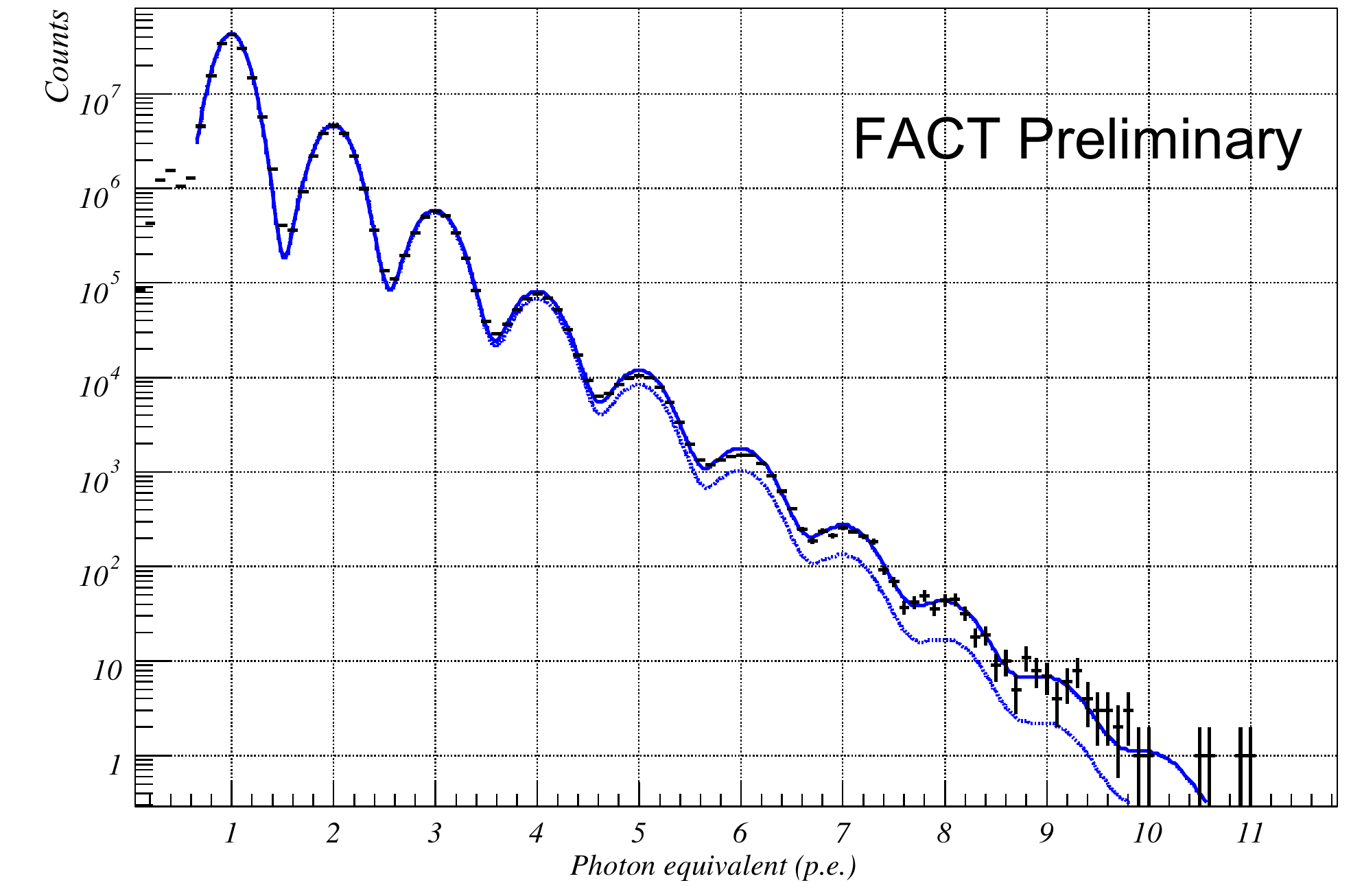}
\caption{Dark count spectrum, for a randomly selected single pixel and a single run (left), and normalized spectrum compiled from all dark count runs and all channels (right). The solid like is a fit with the function explained in the text. For comparison, the dashed line shows a fit with the coefficient \(r\) fixed to zero.}
\label{fig:spectrum}
\end{center}
\end{figure*}
\begin{figure*}[t]
\begin{center}
 \includegraphics*[width=0.18\textwidth,angle=0,clip]{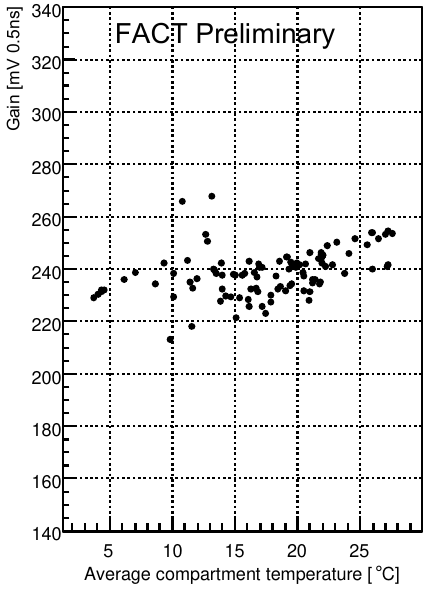}
 \includegraphics*[width=0.38\textwidth,angle=0,clip]{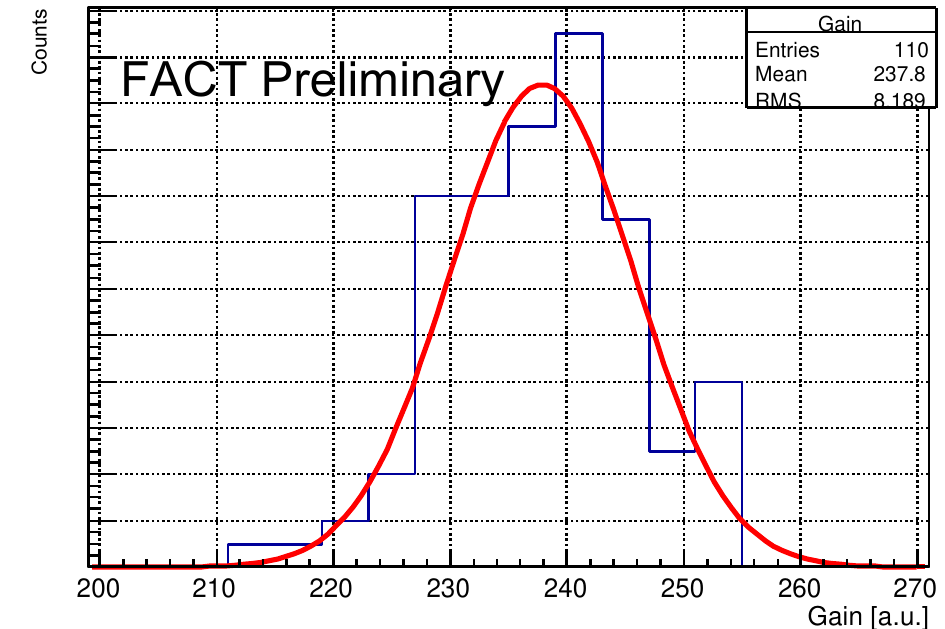}
 \includegraphics*[width=0.38\textwidth,angle=0,clip]{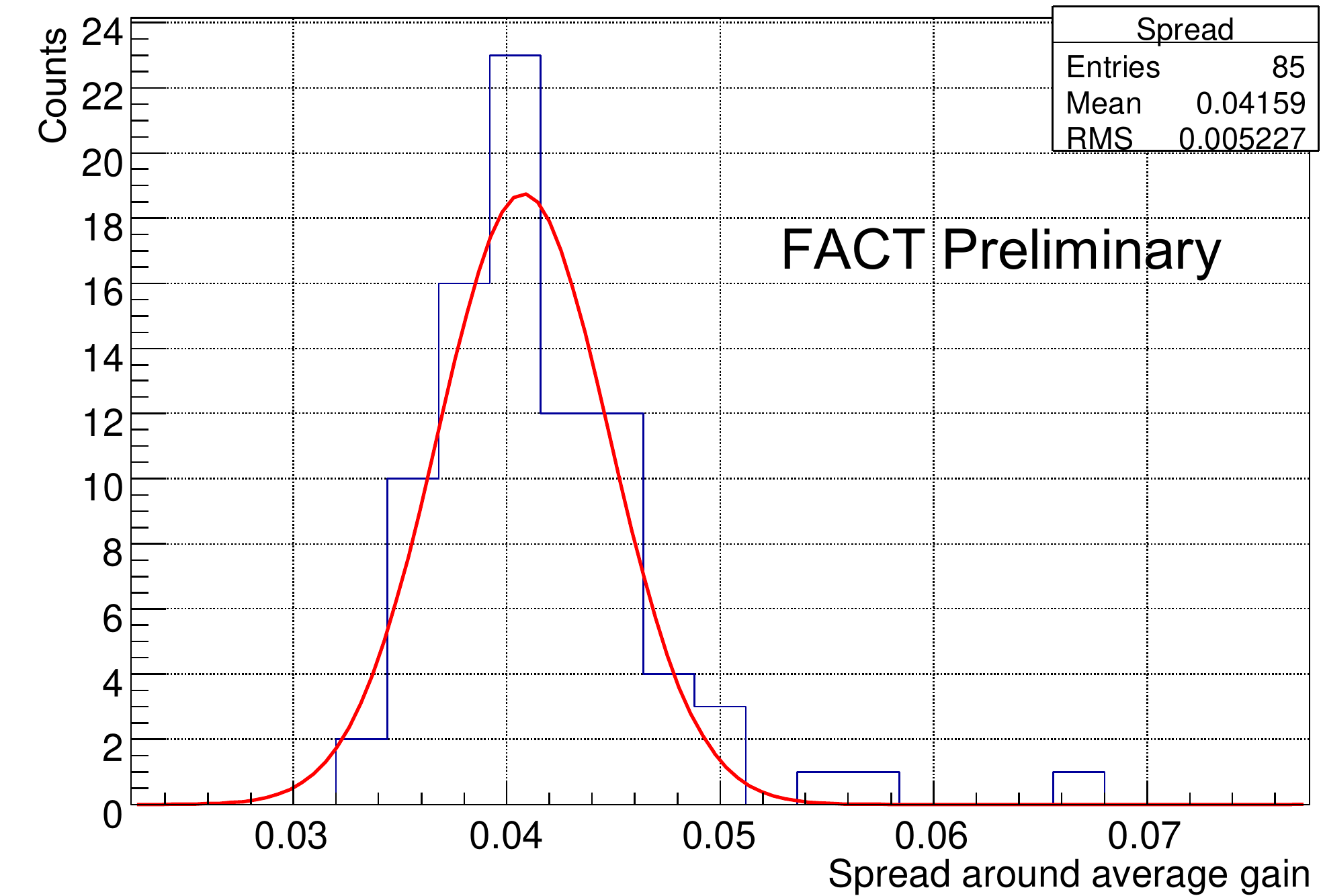}
\caption{Extracted gain (left) versus average compartment temperature
(left) and the corresponding projection (middle) with a fitted
Gaussian. 
The right plot shows the corresponding average
difference between the gain extracted from the individual channels and the
gain obtained from the sum spectrum fitted with a Gaussian.}
\label{fig:gain}
\end{center}
\end{figure*}
\begin{figure}[t]
\begin{center}
 \includegraphics*[width=0.31\textwidth,angle=0,clip]{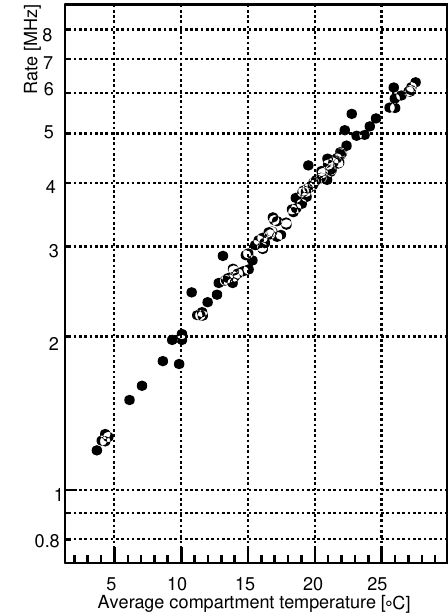}
\caption{An estimate of the dark-count rate obtained from the
integration of the normalized sum-spectra versus average compartment
temperature, extraction efficiency and dead-time are neglected. The correlation
fits very well with the data-sheet.} 
\label{fig:DarkCounts}
\end{center}
\end{figure}

In Fig.~\ref{fig:TempDist}, the distribution of the light-pulser amplitude is
shown. The sigma of the Gauss fit yields \(\pm 6\%\) of the mean
value.

\subsection{Direct gain measurement}

Due to the high precision of the charge released in breakdowns of single 
G-APD cells, they feature 
excellent resolution for single signals (single-\pe). This allows the extraction
of their properties from their dark count spectrum. For this, randomly 
triggered data is recorded with closed lid. The comparably low
rate ensures that the probability for 
two coinciding dark count signals is reasonably low 
simplifying their identification and analysis.

\paragraph{Gain extraction}

The quality of the gain measurement relies entirely on the quality of the
single-\pe{} spectrum, i.e.\ the recognition
and extraction of pulses induced from a single primary breakdown.
The gain is determined from the distance of higher order peaks to
avoid a bias from a possible baseline shift, although the
baseline for each channel is determined and corrected.
For this study, the dark count spectrum of all 1440 channels has been
investigated individually. 
For each measurement a run with 3000 randomly triggered events was taken.

To extract the pulses, a sliding average of ten samples is computed
from the data. This improves the signal-to-noise ratio and eliminates some
occasional noise with small amplitudes and a fixed frequency.
To identify pulses, the remaining samples are scanned for a leading edge
defined by a threshold-crossing and the constrain that four samples
before and four samples after that point the signal must still be below
and above this threshold, respectively.
From this
point, the local maximum is determined within the following 30 samples
and the point at which the leading edge reaches 50\% of the local
maximum, called {\it arrival time of the pulse} hereafter. If the
distance between the maximum and the arrival time exceeds 7\,ns the
pulse is discarded. From the arrival time position onwards, 30 samples
of the raw-signal are integrated. Since the arrival time is the point
of the steepest slope, the integration starting at this point yields the
most stable results, which was confirmed by simple cross-checks.

\paragraph{Fit function}

The resulting distribution is a superposition of the distribution of
the single-\pe{} peak and higher order peaks resulting from optical
crosstalk. Each distribution is believed to be Gaussian with a width
\(\sigma_{n}\) being compiled from a constant noise component
\(\sigma_{el}\), and a noise originating from the fluctuations of the
amplitude of a single avalanche, \(\sigma_{pe}\). 
The ratio between the number of events represented by two
consecutive peaks is described by the probability \(p\), the probability
that an avalanche induces exactly one other avalanche in another micro-cell.

In the obtained spectra, a deviation from this strict exponential
behavior is observed. 
Since this study does not aim at the precise determination of other
parameters than the gain, this effect is not discussed in
details. 

\paragraph{Fit procedure}

To determine reasonable start values for fitting the spectra of all
channels individually, a combined spectrum of all
pixels is filled and fitted for each run.
With the baseline and
gain value determined for each channel, the individual spectra are
normalized and compiled into a sum-spectrum. Fig.~\ref{fig:spectrum} (right)
shows the sum of these spectra of all available data and a fit with the
function given above. The very good agreement of the individual
spectra after normalization is evident.

\paragraph{Result}

From the primary sum-spectrum, the average gain of all channels
is determined. The dependence of the extracted gain versus the
average sensor temperature is shown in Fig.~\ref{fig:gain} (left).
A small
remaining dependence of the gain on temperature is visible which is of
the order of 5\% per 20\,K. This is an effect of a non-ideal
temperature adaption coefficient in the control software which has
intentionally not been touched since the deployment of the camera to
ensure consistency of the obtained data during a reasonably long
period. The center plot shows the distribution of the measured gain
values. The fitted Gaussian yields a width of the distribution of about
\(\pm 3\%\) which is well within the specification requiring a gain
stability of \(5\%\). The right distribution corresponds to
the average root-mean square of the difference between the gain
obtained from the single pixel fit and the average gain obtained from
the fit of the combined spectrum of all pixels. It can be interpreted
as the  typical gain spread of pixels within the camera. The fitted
Gaussian yields a mean of slightly more than 4\%.

As a cross-check, the dark count rate can be estimated from the spectra
and plotted versus temperature, see Fig.~\ref{fig:DarkCounts}. This
measurement is not a precise measurement of the dark count rate because
dead-time and efficiency effects are neglected. 
Nevertheless, the obtained dark count rate fits very well the
data-sheet values.

\subsection{Ratescans}

Another method to access the gain are so called ratescans. Ratscans use
the cosmic ray spectrum for an indirect measurement. They measure
the dependence of the total trigger rate of the system on the trigger
threshold settings.

The trigger of the  camera comprises two stages. In the first
stage, the signals of nine channels are summed.
This sum-signal is clipped to suppress effects of baseline fluctuations in
case of high noise conditions. A comparator converts this signal into a
a digital signal. Four of these signals are again summed and
discriminated slightly below the threshold for a single input
signal. This second step mainly reduces fake triggers from noise of the
electronics itself by suppressing too short trigger signals.

Changing the comparator thresholds, the dependence of the total trigger
rate on the threshold settings can be investigated. For low thresholds
the rate is dominated by noise and photons from the diffuse night-sky
background. For high thresholds, random triggers by artificial
coincidences are suppressed, and only simultaneously arriving
photons from cosmic-ray induced showers will trigger the system.

\begin{figure}[t]
\begin{center}
 \includegraphics*[width=0.48\textwidth,angle=0,trim=0 0 1.cm 0.4cm,clip]{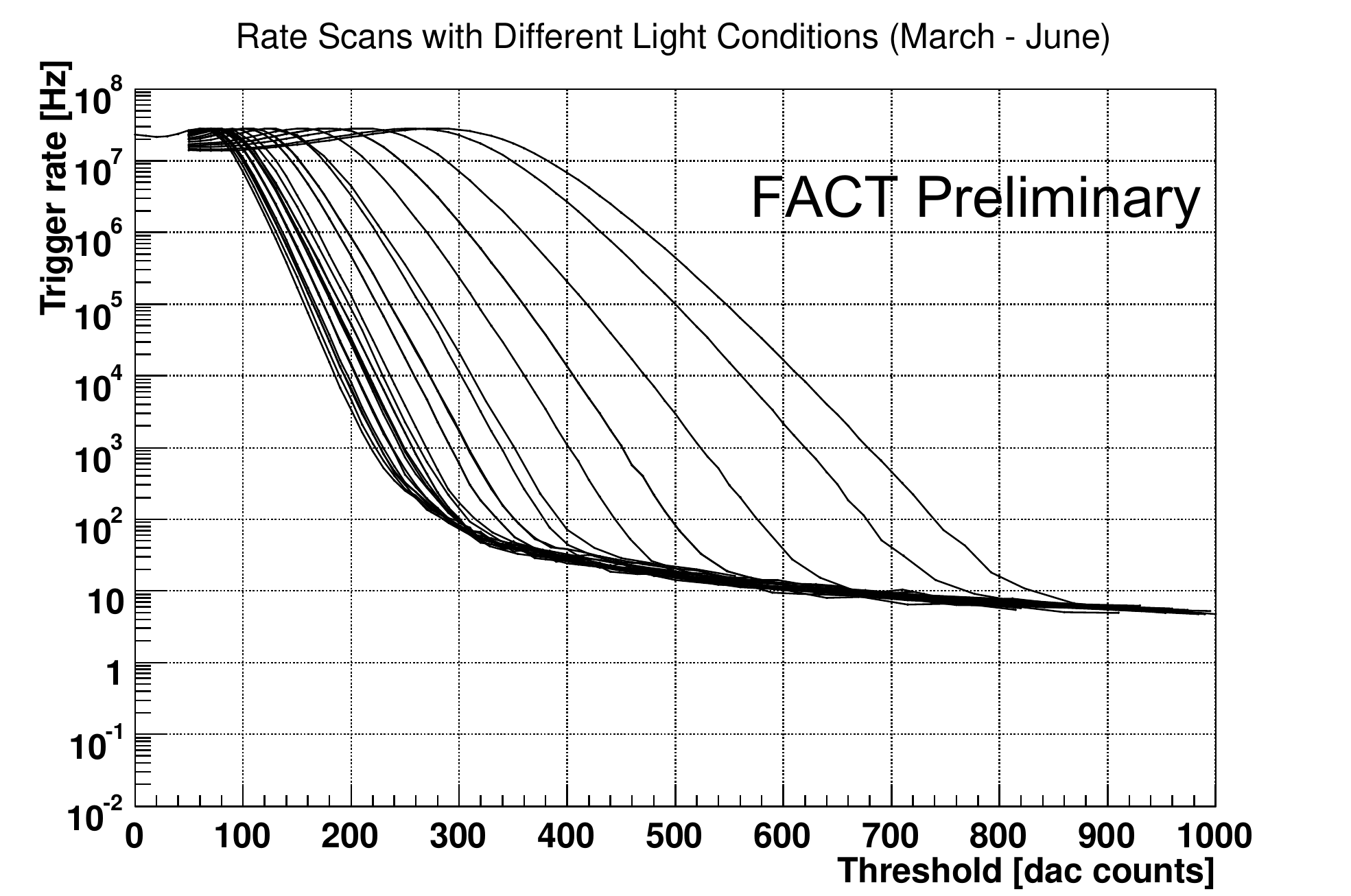}
\caption{Ratescans taken with the feedback system enabled and
different light conditions at nominal voltage during March 2012 to June
2012.}
\label{fig:Ratescans0V}
\end{center}
\end{figure}

\paragraph{Results}

For rates above 10\,MHz, the system saturates.
Below, the trigger rate is dominated by random
coincidences of night-sky background photons turning
into triggers caused by coincident photons from cosmic-ray induced
showers.

To prove the stability of the gain over a longer time-period and its
independence of the ambient light condition, several ratescans have been
taken between March and July 2012 with changing light conditions.
Fig.~\ref{fig:Ratescans0V} shows 26 ratescans taken with light
conditions ranging from dark night to almost full moon
(\(\approx 90\%\)). It is visible that the shower induced part of the
curves is independent of the ambient light conditions corresponding to a stable
gain.

More on measured current, light condition,
trigger threshold and rate can be found in~\cite{bib:threshold,bib:ratescans,bib:moon}.

\section{Conclusion}

The camera of the FACT telescope is now operated since 20 months
and has proven the applicability of silicon photo sensors
in real-life focal plane installations, in particular, in Cherenkov telescopes.

To achieve a stable operation, of the applied Geiger-mode avalanche 
photo diodes, a feedback system adapting the applied voltage
according to temperature readings and the measured current has been developed.

Three different methods to measure the gain
of the system directly or indirectly were applied. They show consistent
results on gain-stability. The presented results are limited
by the calibration procedure of the bias voltage system, which is currently
improved. The most precise and direct method, the extraction
of the dark count spectrum, has shown a long-term stability over several
months under changing temperature conditions at the few percent level.
The measurement of the amplitude of an external light-pulser signal
has proven the stability independent of the background light level also
at the few percent level. 

The implemented feedback system renders the need for an external calibration
device obsolete, which is a big advantage for further Cherenkov telescope
projects saving a lot of development time and costs.

A detailed description of the telescope and camera hardware and software can be found
in~\cite{bib:status,bib:design}. A more detailed discussion of the stability will be
available soon in~\cite{bib:feedback}.

\footnotesize{\paragraph{Acknowledgment}{The important contributions
from ETH Zurich grants ETH-10.08-2 and ETH-27.12-1 as well as the
funding by the German BMBF (Verbundforschung Astro- und
Astroteilchenphysik) are gratefully acknowledged. We are thankful for
the very valuable contributions from E.\ Lorenz, D.\ Renker and G.\
Viertel during the early phase of the project We thank the Instituto de
Astrofisica de Canarias allowing us to operate the telescope at the
Observatorio Roque de los Muchachos in La Palma, and the
Max-Planck-Institut f\"ur Physik for providing us with the mount of the
former HEGRA CT\,3 telescope, and the MAGIC collaboration for their
support. We also thank the group of Marinella Tose from the College of
Engineering and Technology at Western Mindanao State University,
Philippines, for providing us with the scheduling web-interface.}}

\end{document}